# Tracking Coupled Granular Temperature and Configurational Entropy Dynamics in Granular Materials via Dielectric Spectroscopy

**Sophia G. Krastana (Σοφία Γ. Κραστάνα), Anthony N. Papathanassiou (Αντώνιος Ν. Παπαθανασίου)**

National and Kapodistrian University of Athens, Physics Department, Condensed matter Physics Section, Panepistimiopolis, 15784 Zografos, Athens, Greece

## Abstract

In glass-forming liquids, structural dynamics are closely related to configurational entropy and temperature, with dielectric relaxation time scaling alongside structural relaxation time as described by the Adam-Gibbs (AG) model. Under Edwards's athermal statistical thermodynamics, a modified AG relationship has been proposed for granular matter, provided that granular temperature and configurational entropy are appropriately defined. This study investigates whether the electrical response of granular systems can provide insight into their structural dynamics through thermally activated processes, specifically electric charge hopping and trapping. By progressively reducing the volume of graphite powder to vary its packing fraction, we estimated relative configurational entropy and granular temperature from volumetric data, while evaluating electrical conductivity and capacitance via impedance spectroscopy. We demonstrate that the logarithm of the dielectric relaxation time, derived from complex impedance, exhibits a scaling behavior with granular temperature and configurational entropy across both loose and compact states. Consequently, changes in the complex impedance resulting from packing fraction variations are tuned by granular configuration, consistent with an AG-like relationship for thermal systems. These findings establish dielectric spectroscopy as a viable, non-destructive tool for tracing configurational dynamics in granular matter, analogous to its established use in polymers and glass-forming liquids.

**Corresponding author:** Anthony N. Papathanasiou, E-mail address: antpapa@phys.uoa.gr

## 1 Introduction

Granular materials are assemblies of discrete solid particles, typically exceeding one micron in size, rendering them effectively athermal. Because thermal fluctuations are negligible at this scale, the dynamic behavior of these systems is driven primarily by external macroscopic forces, such as gravity, shear, or mechanical vibration [1–3]. The macroscopic behavior of granular matter is governed by inter-particle forces that are fundamentally frictional and non-conservative. Due to friction at grain-to-grain contacts, kinetic energy within granular systems is continuously dissipated as heat, causing these materials to remain far from equilibrium and to become arrested in metastable states. Their collective dynamics are therefore driven by external forcing, as internal dissipative mechanisms prevent the system from relaxing to a stable equilibrium state [1, 2]. Granular materials exhibit bimodal characteristics, flowing like fluids under certain conditions while responding like solids when the applied stress exceeds a critical threshold [1].

Supercooled and glass-forming liquids—liquids quenched below their freezing points without undergoing crystallization—represent complex, far-from-equilibrium thermodynamic systems. Within these metastable states, thermal fluctuations continue to drive microscopic degrees of freedom, inducing atomic or molecular dynamics across translational, vibrational, and rotational modes [1–6]. During transitions between non-equilibrium states, materials exhibit significant shifts in their characteristic structural relaxation time defined as the timescale required for atomic groups to undergo rearrangement and lose their local configurational order [4, 5]. These underlying structural dynamics can be robustly described using the Adam-Gibbs (AG) framework, which provides a critical theoretical link between the macroscopic thermodynamic state of the system and its microscopic relaxation behavior [7].

According to this model, the kinetics of structural relaxation are governed by configurational entropy, which establishes the characteristic scale of cooperatively rearranging regions (CRRs) [7]. To achieve localized configurational changes, atoms or molecules within these regions must undergo collective reorganization. Thus, the temperature dependence of the structural relaxation time reflects an energetic competition between the macroscopic energy barrier of this cooperative atomic rearrangement and the configurational entropy of the system and $T\Delta S_c(T)$, where $T$ denotes the absolute temperature and $\Delta S_c(T)$ is the difference between the configurational entropy $S_c(T)$ and a reference value $S_0$. As configurational entropy diminishes, molecular mobility decelerates, significantly increasing the time necessary for the system to explore new metastable configurations [4, 7, 8]. The dramatic deceleration of relaxation in glass-forming systems reflects a depletion of accessible configurations, limiting the exploration of microstates [8]. To rationalize this phenomenon, the Adam-Gibbs (AG) model provides a theoretical framework, linking thermodynamic constraints to the emergence of cooperative dynamics in deeply supercooled materials [4, 5].

In both glass-forming liquids and granular media, the structural relaxation time characterizes the fundamental timescale for constituent particles to rearrange within the bulk. This timescale increases significantly as these systems approach kinetic arrest either through cooling toward the glass transition or through increasing packing density toward the jamming transition [9–11]. In supercooled and glass-forming liquids, microstate transitions are driven by thermal fluctuations, with the Adam-Gibbs (AG) theory linking the system's thermodynamics to its relaxation dynamics [7]. Conversely, in athermal systems such as granular materials, thermal energy is insufficient to induce structural rearrangements. Instead, these systems rely primarily on external mechanical excitations to transition between microstates [1–3].

Within the framework of Edwards' statistical mechanics, the static configurations of a system are described using concepts analogous to those of classical Boltzmann thermodynamics, namely compactivity and configurational (or granular) temperature. Granular temperature is defined thermodynamically as the rate of entropy change upon the absorption of mechanical energy at constant volume, where the configurational entropy is related to the physical tessellation of the specimen's volume [12–14]. Accordingly, a modified Adam-Gibbs relationship for the structural relaxation of granular systems has been proposed in the literature [17, 18].

In supercooled liquids and polymers, the glass transition is intrinsically linked to the α-relaxation, a dynamic process that can be effectively probed via Broadband Dielectric Spectroscopy (BDS) [19,20]. In these systems, structural relaxation upon approaching the glass transition temperature is evidenced by distinct changes in the dielectric relaxation time. Inspired by the established correlation between structural and dielectric relaxation times in conventional glass formers, this study investigates whether a similar relationship exists in granular matter, specifically graphite powder, under varying packing fractions. Modifying the packing fraction induces spatial rearrangement of the granules, optimizing their distribution within the available volume. As the packing fraction increases, the charge percolation network is enhanced and the morphology of inter-grain pores is altered, resulting in measurable changes in electrical conductivity, capacitance, and dielectric relaxation time. By comparing the dielectric relaxation response in relation to packing fraction and the corresponding thermodynamic quantities, we investigate whether the observed electrical dynamics are consistent with entropic variations and, consequently, evaluate the applicability of Adam-Gibbs (AG) model to these materials.

## 2 Materials and methods

### 2.1 Experimental setup

Graphite powder (Sigma-Aldrich, CAS No. 7782-42-5) with a particle size of ≤ 20 μm was used as the granular material. The powder was stored in a sealed, desiccated container and used as received without further pretreatment. The dielectric properties of the powder samples were evaluated using a Marconi TJI 55 C/I capacitor-type dielectric loss tester. This instrument features two parallel, horizontal, disk-shaped electrodes; one electrode is movable with micrometer precision, allowing the inter-electrode spacing to be accurately measured via an integrated micrometer.

A weighed quantity of graphite powder was placed into a custom-assembled, cup-shaped sample holder comprising a hollow Teflon cylinder with a flat tin disk electrode rigidly mounted at its base (Figure 1). This assembly was positioned between the parallel electrodes of a Marconi TJI 55 C/I capacitor-type dielectric loss tester. The tin base electrode was concentrically aligned with the tester's lower horizontal electrode, and silver paste was applied to the interface to ensure optimal electrical contact.

The inner diameter of the Teflon cylinder precisely matches that of the upper moving electrode. This geometry ensured proper powder confinement while allowing unrestricted downward movement of the upper electrode onto the sample surface. Additionally, the upper electrode remained unsealed to permit the escape of trapped air during the compression of the granular bed. By systematically decreasing the inter-electrode distance—thereby reducing the volume of the confined powder—simultaneous, high-accuracy measurements of volume and dielectric properties were performed on a fixed mass of graphite.

### 2.2 Volumetric measurements

The total volume of the granular material—comprising both the solid grains and the interstitial pore space—was determined experimentally by measuring the inter-electrode distance with a micrometer. The packing fraction is defined as the ratio of the volume $V_S$ of a quantity of solid grains to the entire macroscopic volume $V$ occupied:

$$\varphi \equiv \frac{V_s}{V}. \quad (1)$$

If $m_s$ denotes the mass of solid granules of density $\rho_s$, the effective density of the granular system is defined as $\rho \equiv m_s/V$. The definition establishes a direct relationship between $\varphi$ and $\rho$: $\varphi = \rho/\rho_s$.

### 2.3 Broadband Dielectric Spectroscopy measurements

A Marconi TJI 55 C/I dielectric loss tester was connected to a Zurich Instruments MFIA lock-in amplifier via two 2-m coaxial cables for two-terminal measurements. Compensation procedures implemented in the LabOne software (Zurich Instruments) were used to perform user-defined calibrations to correct parasitic wiring resistance and capacitance across the entire frequency range. The MFIA impedance analyzer applied an AC signal to the sample while sweeping the frequency from 1 Hz to $10^5$ Hz. Upon each sequential reduction in sample volume, the complex impedance was measured as a function of frequency for individual values of the packing fraction.

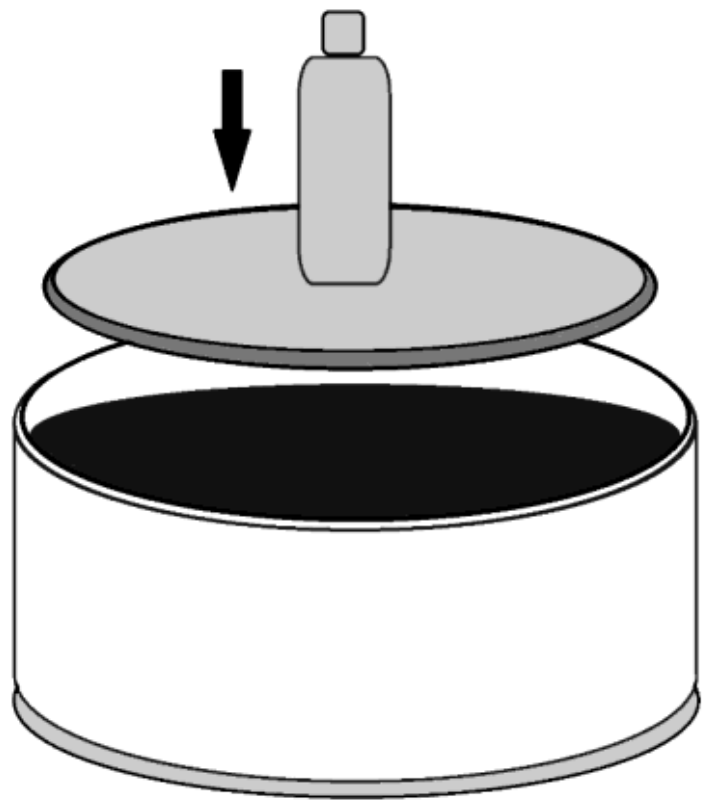

**Figure 1:** Schematic of the powder holder assembly mounted on the lower horizontal electrode of a Marconi TJI 55 C/I dielectric loss tester. A hollow Teflon cylinder (white) is securely sealed at its base by a disc-shaped tin electrode. The upper electrode translates vertically within the Teflon tube to confine and gradually compress the powder sample. Micrometer adjustment of the inter-electrode distance permits accurate determination of the total volume of granular matter.

## 3 Theoretical backgrounds

### 3.1 Thermodynamic approach within Edwards' theory

In atomic systems, thermal energy drives the continuous transitions between various microstates that constitute a given macrostate. In granular systems, however, thermal energy is negligible compared to the mechanical energy required to displace individual grains. Thermal energy is insufficient to induce significant grain rearrangements. Consequently, granular assemblies remain trapped in static, effectively frozen microstates; transitions between these states require external mechanical perturbations rather than thermal fluctuations, provided temperature remains constant. In the absence of mechanical triggering, changes in temperature may induce thermal expansion of the individual grains, potentially alter the distribution of intergranular forces, and cause localized structural rearrangements. To isolate the effects of mechanical compression, the experiments in the present study were conducted at a constant room temperature. Successive static macroscopic states in dense granular matter are attained by compressing a confined powder under isothermal conditions, thereby reducing its total volume. The change in entropy is attributed to changes in the configurational entropy $s_c$, which is proportional to the logarithm of the number of granular microstates corresponding to the macroscopic energy of the system. In Boltzmann's statistical mechanics of ergodic systems, the proportionality constant is Boltzmann's constant, $k_B$. In athermal granular systems, a different constant $\lambda$ (in $\mathrm{J/K}$ units) is introduced. As will be shown below, the unknown proportionality constant $\lambda$ cancels out in our theoretical analysis. A reduced configurational entropy is defined herein as $S_c \equiv s_c/\lambda$. $S_c$ is dimensionless, while $s_c$ is measured in $\frac{\mathrm{J}}{\mathrm{K}}$. A configurational (sometimes termed granular) temperature $T_c$ of a granular medium containing $N$ grains is defined as:

$$(\lambda T_c)^{-1} = \left(\frac{\partial s_c}{\partial U_C}\right)_N. \tag{2}$$

where $U_C$ denotes the internal energy of the granular system. Within the statistical mechanics framework proposed by Edwards for athermal granular media (encompassing both dissipative and non-dissipative systems), compactivity is formally defined as:

$$X^{-1} = \left(\frac{\partial S_c}{\partial V}\right)_N. \tag{3}$$

Compactivity is a measure of the capacity of a granular system containing $N$ grains to be compressed [1]. The product $T_c \Delta s_c$ can be written as:

$$T_c \Delta s_c = \lambda T_c \Delta S_c = \left(\frac{\partial U_C}{\partial S_c}\right)_N \Delta S_c = \left(\frac{\partial U_C}{\partial V}\right)_N \left(\frac{\partial V}{\partial S_c}\right)_N \Delta S_c = \left(\frac{\partial U_C}{\partial s_c}\right)_N \Delta S_c = \left(\frac{\partial U_C}{\partial V}\right)_N X \, \Delta S_c. \tag{4}$$

In the product $T_c \Delta s_c$, the unknown proportionality constant $\lambda$ cancels out. The chain yielding Eq. (4) is valid if $U_C$, $S_c$ and $V$ are state variables of the same ensemble and that the partial derivatives are evaluated under consistent thermodynamic constraints.

**Determination of the term $\left(\frac{\partial U_C}{\partial V}\right)_N$.** The internal energy $U_C$ of a granular system comprises the translational, rotational, and vibrational kinetic energies and the potential energies of the individual grains. In this experiment, the kinetic energy components remain constant. However, the potential energy $U$ of the confined granular matter changes during compression due to the reduction in the specimen's height. Let us consider a system of hard spheres with a total mass confined within a cylindrical container of cross-sectional area $A$. These spheres constitute a cylindrical column of height $h$ (Figure 1) in a uniform gravitational field with acceleration $g$. A reduction in the height of the cylindrical granular sample by $\mathrm{d}h$ (upon application of an external compressive force normal to the upper electrode) leads to a decrease in the total volume by$\mathrm{d}V$. The change in the potential energy of the system is [23]:

$$\mathrm{d}U = \Lambda m_s g \, \mathrm{d}h = \Lambda m_s g A^{-1} \, \mathrm{d}V. \quad (5)$$

The parameter $\Lambda$ can be set equal to ½ based on some assumptions, such as that the granular system consists of ordered identical hard spherical grains, whose internal energy is exclusively potential energy in the presence of a homogeneous gravitational field. Using Eq. (1), Eq. (5) can be written as:

$$\mathrm{d}U_C\left(\frac{1}{\varphi}\right) = \Lambda m_s g A^{-1} \, V_s \mathrm{d}\left(\frac{1}{\varphi}\right) = -\Lambda \frac{m_s^2 g}{A\rho_s}\left(\frac{1}{\varphi^2}\right) \mathrm{d}(\varphi). \quad (6)$$

In our experiments, the descent of the center of mass of the cylindrical column of granular matter upon compression yields:

$$\frac{\mathrm{d}U_C(\varphi)}{\mathrm{d}\varphi} = -\Lambda \frac{m_s^2 g}{A\rho_s}\left(\frac{1}{\varphi^2}\right). \quad (7)$$

As mentioned above, Eq. (7) was derived assuming that a column of identical frictionless spheres is simulated as a uniform column [23]. It should therefore be used critically as a crude approximation when applied to realistic (frictional, dissipative history-dependent compressed) systems.

**Determination of the term $XS_c$ from volumetric data.** Aste and Di Matteo [24] studied the distribution of volume fluctuations in experiments and numerical simulations of equal-sized sphere packings prepared using different techniques. It was observed that the distributions of local (Voronoï cells) and global (entire specimen) volumes are well described bay shifted and rescaled Gamma distribution, known as "k-gamma distribution." Such agreement is observed over a broad range of packing fractions, and it is observed for several distinct systems [24].

Oquendo et al. [17], building on the proposal by Aste and Di Matteo [24] to measure Edwards's compactivity from the volume distribution of Voronoï or Delaunay tessellations ,

derived an equation of state relating the compactivity to the packing fraction by assuming that the total volume can be divided into elementary cells of fixed minimum volume The validity of this approach and its underlying assumption was confirmed by molecular dynamics simulations for volumetric behavior of three-dimensional ensembles of monodisperse spheres at both the limit state of isotropic compression and the limit state of shear, over a broad range of the sliding and rolling friction coefficients. A compactivity—packing fraction equation without any free adjustable free parameters were expressed in terms of the volumetric characteristics of the system [17].

$$X(\varphi) = A'\left[\frac{1}{\varphi} - \frac{V_{min}^{Voro}}{v_{grain}}\right], \quad (8)$$

where $V_{min}^{Voro}$ is the minimum volume of a 3D Voronoi cell corresponding to a grain of volume $v_{grain}$. Geometrical calculations for randomly packed three-dimensional systems consisting of identical hard spheres of radius $d$, indicate that $\frac{V_{min}^{Voro}}{v_{grain}} \cong 1.3250$ is a constant [17] and $A' = \frac{X_{RCP}}{\left(\frac{1}{\varphi} - \frac{V_{min}^{Voro}}{v_{grain}}\right)} \cong 0.04421d^3$ [25].

Integrating Eq. (3), the configurational entropy change associated with a change in the total volume from $V_1$ to $V_2$ is obtained:

$$\Delta S_c = \int_{V_1}^{V_2} X^{-1}(V)\mathrm{d}V = \int_{\varphi_1}^{\varphi_2} \frac{V_s}{X\left(\frac{1}{\varphi}\right)} \mathrm{d}\left(\frac{1}{\varphi}\right). \quad (9)$$

An analytical expression for $X\left(\frac{1}{\varphi}\right)$ is given by Eq. (8). Hence, Eq. (9) reads:

$$\Delta S_c = \int_{\varphi_1}^{\varphi_2} \frac{V_s}{A'\left[\frac{1}{\varphi} - \frac{V_{min}^{Voro}}{v_{grain}}\right]} \mathrm{d}\left(\frac{1}{\varphi}\right) = \frac{V_s}{A'}\int_{\varphi_1}^{\varphi_2} \frac{1}{\left[\frac{1}{\varphi} - \frac{V_{min}^{Voro}}{v_{grain}}\right]} \mathrm{d}\left(\frac{1}{\varphi}\right) = \frac{V_s}{A'}\left[ln\left(\frac{\frac{1}{\varphi_2} - \frac{V_{min}^{Voro}}{v_{grain}}}{\frac{1}{\varphi_1} - \frac{V_{min}^{Voro}}{v_{grain}}}\right)\right]. \quad (10)$$

To determine entropy changes $\Delta S$, a reference value $\varphi_1$ corresponding to a random close packing (RCP) state can be chosen. RCP is identified with the densest packing for which no crystalline order is present [9–11]. For randomly close packed spheres, a value of $\varphi_{RCP} \cong 0.64$ is commonly regarded as representative, supported by first-principles statistical mechanics calculations [26]. Subsequently, the entropy change $\Delta S_c$ can be calculated from Eq. (10), by setting $\varphi_1 \cong 0.64$. Two key considerations regarding the estimation of configurational entropy. First, while existing compactivity and entropy models rely on simplifying assumptions about particle size and morphology—which may introduce systematic errors into absolute calculations, these errors cancel out when computing relative entropy differences. Second, because the physical significance lies in relative entropy changes rather than absolute values, a baseline reference packing fraction $\phi_{\rm ref}$ can be chosen arbitrarily, for example, as the initial packing fraction in a sequence of compression steps.

### 3.2 Modified Adam-Gibbs (AG) model for granular matter

The Adam-Gibbs (AG) model provides a foundational relationship between the thermodynamic states and dynamic relaxation behavior of liquids and ultra viscous glass-forming systems, in terms of their configurational entropy [7]. The structural relaxation time $\tau_s$ of a supercooled liquid is related to the configurational entropy $\Delta S_C$ which determines the size of cooperatively rearranging regions:

$$\tau_S = \tau_0 exp\left(\frac{A''}{T\Delta S_C}\right), \tag{11}$$

where $\tau_0$ and $A''$ are constants, and $\Delta S_C$ is the configurational entropy, which can be defined as the difference between the entropy of the disordered system and that of its crystalline state. In the AG model, the transition from one local equilibrium minimum to another is *thermally activated* and involves cooperative atomic or molecular conformational motions. Consequently, the activation energy scales inversely with the configurational entropy $\Delta S_C(T)$ Recent studies on granular systems [27] have suggested that structural relaxation dynamics in athermal systems can be described by a modified Adam-Gibbs (AG) relation, by replacing the term $T\Delta S_C$ appearing in Eq. (11) with $T_c\Delta s_c$,

$$\tau_S = \tau_0 exp\left(\frac{A''}{T_c\Delta s_c}\right). \tag{12}$$

The validity of a modified AG relationship (Eq. (12)) in tapped granular packings was recently confirmed experimentally [27]. Changes in the configurational entropy of granular matter are driven primarily by external mechanical stress rather than temperature. In our experiments, subjecting the confined powder to compressive stress results in a reduction in its total volume. This compaction process involves localized spatial rearrangements of the individual grains, leading the system towards a quasi-equilibrium state characterized by a reduced macroscopic volume. The structural relaxation time $\tau_s$ characterizes the stress-induced transitions between states and the associated changes in $\Delta S_C$. In the present work, a moving electrode applies this external mechanical stress to the upper boundary of the confined sample. We assume that Eq. (12) can be used to relate structural relaxation time $\tau_S$ to the granular temperature $T_c$ and the configurational entropy term $\Delta s_C$.

### C. Relationship between structural relaxation time $\tau_S$ and dielectric relaxation time $\tau$.

Dielectric relaxation phenomena are closely related to localized structural dynamics, spatial arrangements, and mechanical rigidity transitions in non-crystalline systems [28].

Complex impedance measurements typically distinguish between macroscopic DC conductivity and spatially localized transport, such as interfacial polarization or dipole rotation. In this work, we analyze conducting graphite powder at different packing fractions. Under an applied electric field, macroscopic DC conduction occurs across a percolative network of physically contacting grains. The efficiency of this transport mechanism depends on the granular coordination number (the mean number of intergranular contacts). Accordingly, the in-phase impedance component, $ReZ$, reflects the DC conduction response and serves as a proxy for the network's physical connectivity. Because the packed powder is a heterogeneous composite of solid grains and air voids, electrical inhomogeneities trap free electrons at the interfaces. This trapping produces capacitive effects that are reflected in the out-of-phase impedance component, $ImZ$, allowing us to relate the measured capacitance to the system's packing density.

External stress alters the physical configuration of the granular matrix, thereby modulating the percolation pathways available for charge transport and redistributing charge trapping centers. Accordingly, the structural evolution of the granular matter mirrors its varying electrical conductivity and capacitive properties. These scaling trends across a range of packing fractions ($\varphi$) are consistent with a proportionality relationship, or at least a close coupling, between dielectric and structural relaxation dynamics. The results support the applicability of the modified Adam-Gibbs relationship to the dielectric response of granular matter and provide an experimental approach for probing its thermodynamics.

## 4 Results and discussion

As illustrated in Figure 1, the upper electrode of the powder sample holder was incrementally displaced toward the stationary lower electrode. This resulted in the uniaxial compression of the granular material and a subsequent reduction in specimen volume. The packing fraction $\varphi$ was then determined using the measured volume of the graphite powder volume at each stage of compression and the density of graphite, $\rho_s = 2260\ \mathrm{kg/m^3}$.

The complex impedance, $Z(f) = \mathrm{Re}Z(f) + i\mathrm{Im}Z(f)$, $i^2 = 1$, of the graphite powder was recorded at different packing fractions using Broadband Dielectric Spectroscopy. The real part, $\mathrm{Re}Z(f)$, is related to electrical conduction in granular media, which results from the interplay between intra-grain and inter-grain transport. These inter-granular pathways involve ohmic conduction at physical interfaces and quantum tunneling through the effective potential barriers of neighboring particles [28]. The imaginary part, $\mathrm{Im}Z(f)$, is associated with electric charge trapping, which arises from the structural and electrical inhomogeneity of the specimen, resulting in capacitive effects.

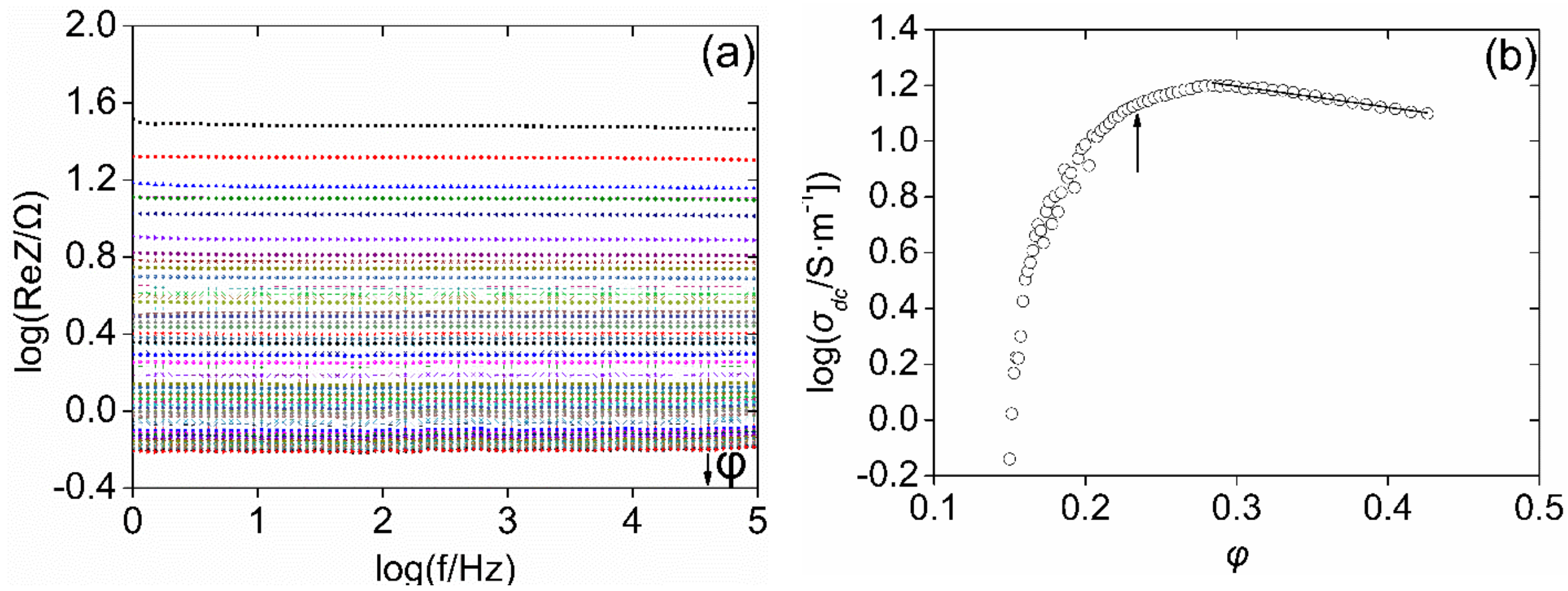

**Figure 2:** (a) $logReZ$ as a function of $logf$. The curves are ordered in ascending $\varphi$ from top to bottom, as indicated by the arrow. (b) $log\ \sigma_{dc}$ as a function of $\varphi$. A straight line fits the high-compacity data points to enable extrapolation to $\varphi = 1$ and comparison with the conductivity of bulk graphite. The arrow indicates the packing fraction value above which the granular medium transitions to a denser phase that resists external mechanical compression.

Figure 2(a) illustrates the logarithm of the real part of the complex impedance ($logReZ$) as a function of frequency ($f$) for various packing fractions ($\varphi$). Across all measured spectra, it remains independent of frequency over the entire range. Minor deviations observed at the high-frequency limit are attributed to transmission line effects. Consequently, the ohmic resistance ($R_{dc}$) is defined as the frequency-independent plateau value of $ReZ$ as $f \rightarrow 0$. Subsequently, the dc conductivity is defined by $\sigma_{dc} = \frac{h}{R_{dc}A}$, where $h$ denotes the height of the cylindrical granular specimen (i.e., the thickness of the disk-shaped sample) and $A$ is the electrode surface area. The wiring resistance in two-terminal mode was $0.15\ \Omega$, corresponding to a transmission-line baseline at $\log[\mathrm{Re}Z(\Omega)] = -0.82$, which is far below the experimental data points presented in Figure 2(a). Since the wiring contribution is weak compared with the measured value of $\mathrm{Re}Z$, its effect on the measured impedance can be considered negligible.

Figure 2(b) presents the dependence of $\log \sigma_{dc}$ on the packing fraction $\varphi$, at which the granular medium transitions into a dense phase that resists further external mechanical compression. In Figure 2(b), a straight line fits the high-compacity data points. By extrapolating the straight line to $\varphi \to 1$, the dc conductivity of bulk graphite is determined as: $\sigma_{dc}(\varphi \to 1) = 4.7\ \mathrm{Sm}^{-1}$.

The real part of the electric permittivity is given by $\mathrm{Re}\varepsilon = \frac{\mathrm{Im}Z}{\omega C_0 |Z|^2}$, where $\omega$ is the angular frequency ($\omega \equiv 2\pi f$), $C_0$ is the capacitance of the empty capacitor, and $|Z| = \sqrt{(\mathrm{Re}Z)^2 + (\mathrm{Im}Z)^2}$. Figure 3 shows a typical plot of $\log\mathrm{Re}\varepsilon$ as a function of $\log f$, for an indicative packing fraction of $\varphi = 0.23$.

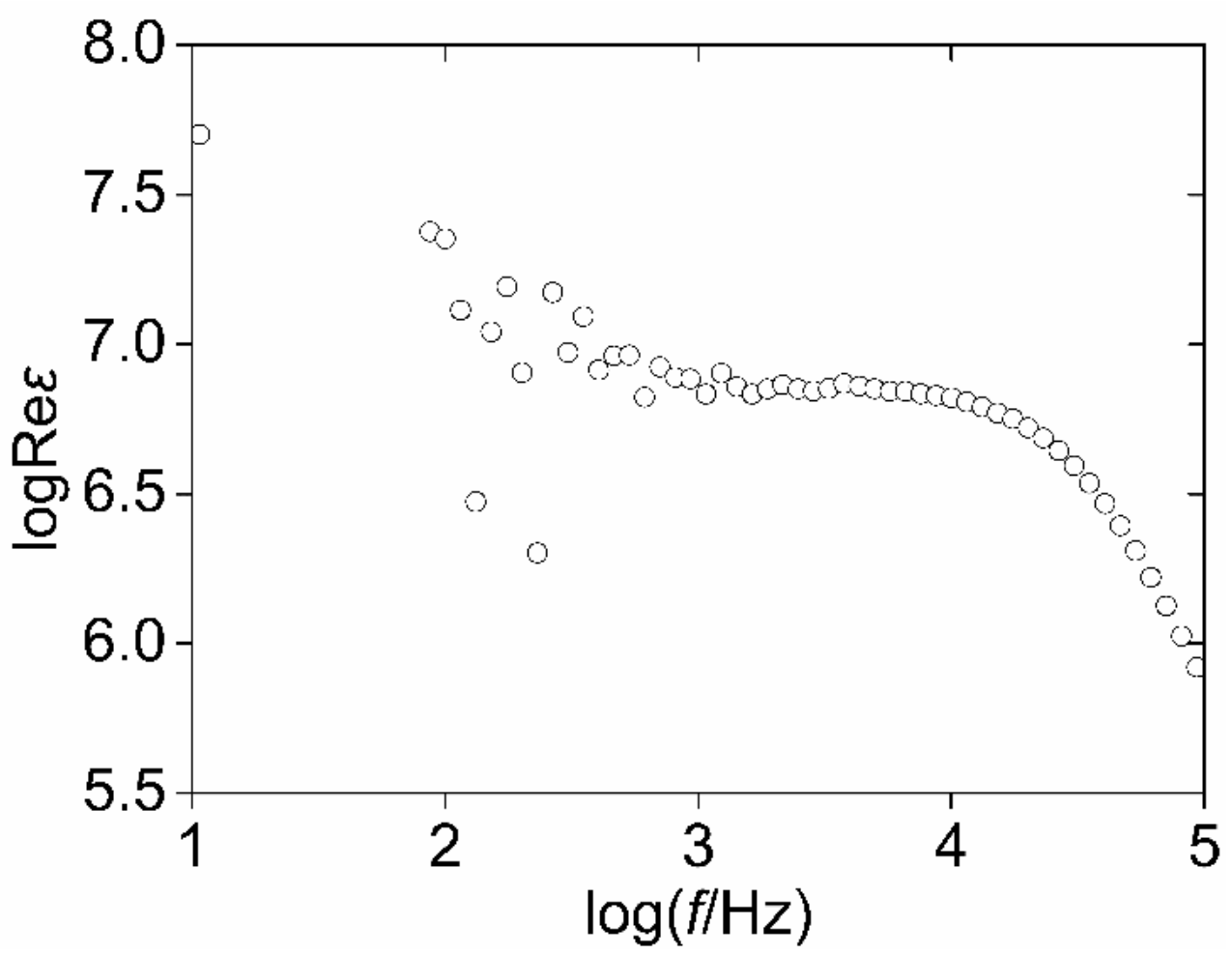

**Figure 3:** $\log\mathrm{Re}\varepsilon$ as a function of $\log f$ for a typical value of packing fraction of $\varphi = 0.23$.

For each packing fraction, the (relative) static permittivity, $\varepsilon_s$, is defined as the value of the real part of the permittivity in the limit of zero frequency (Figure 4). The static capacitance is determined by $C_s = \frac{\varepsilon_s \varepsilon_0 A}{L}$, where $\varepsilon_0$ is the permittivity of free space. The dielectric parameters were characterized as a function of the packing fraction $\varphi$ , which was modulated by incrementally reducing the total volume of the granular sample. From these experimental measurements, direct-current resistance ($R_{dc}$), conductance ($G$), and dc conductivity ($\sigma_{dc}$) were derived. The static (relative) permittivity ($\varepsilon_s$) and static capacitance ($C_s$) were determined. These values facilitated the determination of the effective dielectric relaxation time, defined by the relation $\tau = R_{dc}C_s$.

Figure 4 presents the logarithmic dielectric relaxation time, $\log\tau(\varphi)$, as a function of packing fraction ($\varphi$). By applying the relations defined in Eqs. (5), (6), and (11), the variable $\varphi$ was transformed into the configurational entropy term $T_c \Delta s_c$. This substitution allows $\tau$ to be expressed directly in terms of $T_c \Delta s_c$, resulting in the final scaling plot of $log\tau(T_c \Delta s_c)$. While $\tau$ is obtained from dielectric measurements, $\tau_s$ cannot be determined from Eq. (12) because the e parameters $\tau_0$ and $A''$ remain unknown. However, a linear $\log\tau(T_c \Delta s_c)$ relationship suggested by the combined dielectric and volumetric measurements at various packing fractions would imply that $\tau \propto \tau_s$. The observed scaling is consistent with a proportional relationship, or at least a close coupling, between dielectric and structural relaxation.

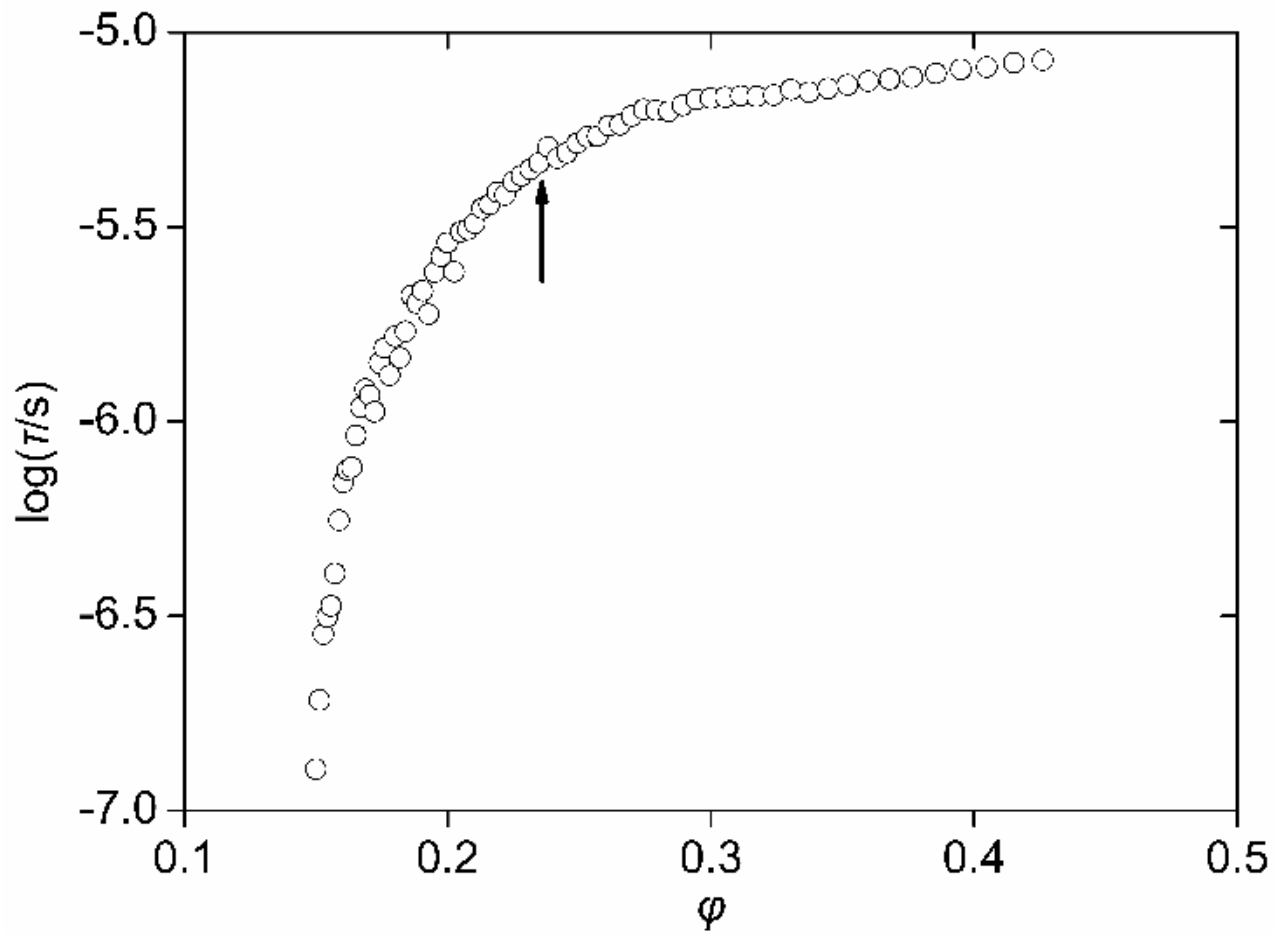

**Figure 4:** $log\,\tau$ as a function of the packing fraction $\varphi$.

Figure 5 presents a plot of $\log(T_c\Delta s_c)^{-1}$ as a function of $\varphi$. $\log(T_c\Delta s_c)^{-1}$ is an increasing function. As the volume decreases, the configurational entropy $\Delta s_c$ is suppressed. This drives the system from a disordered state at low $\varphi$ to a more ordered state at higher $\varphi$, effectively increasing the inverse configurational entropy.

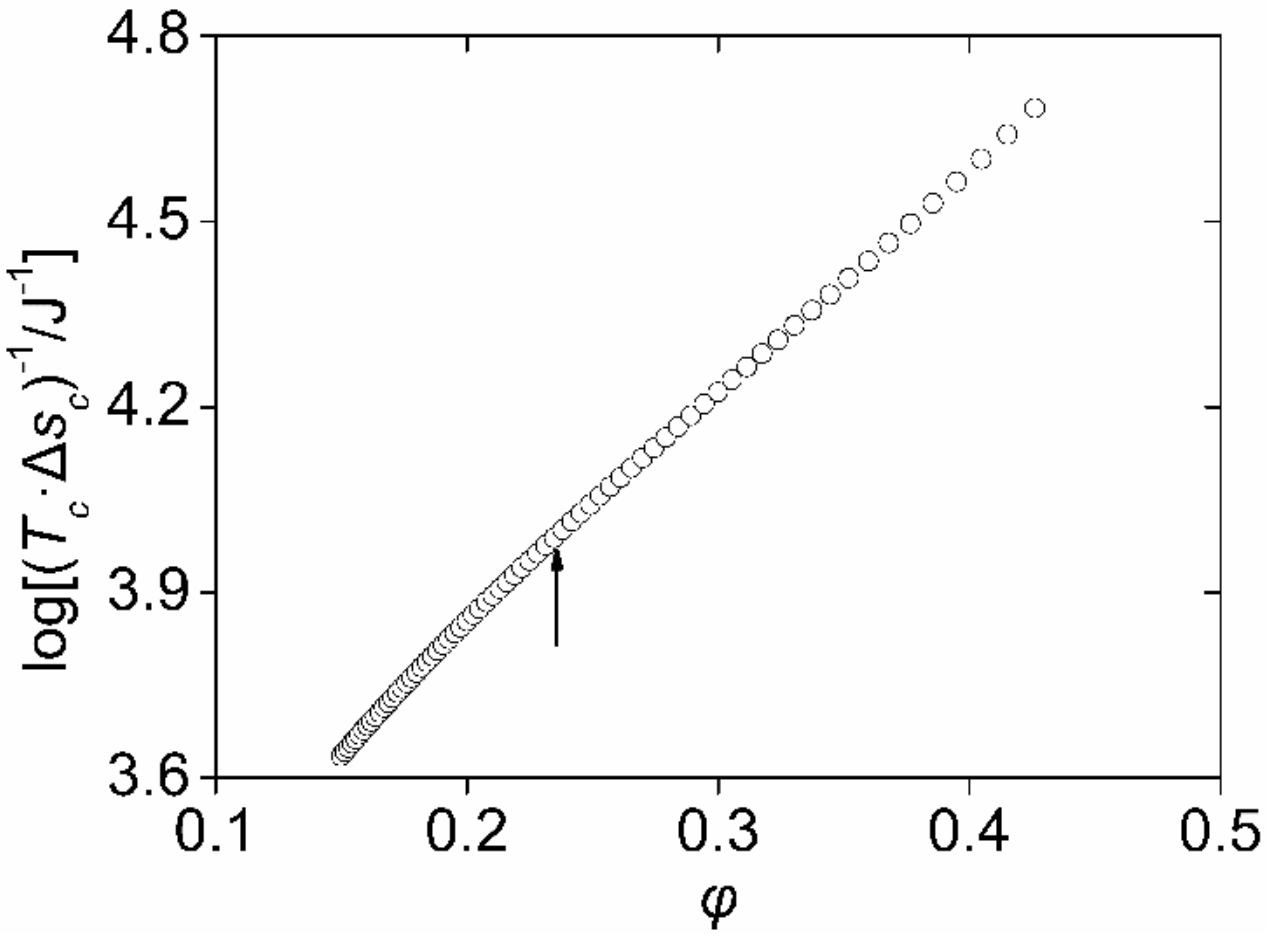

**Figure 5:** $log(T_c\Delta s_c)^{-1}$as a function of the packing fraction.

A plot of $\ln \tau$ as a function of $(T_c\Delta s_c)^{-1}$ is shown in Figure 6. The data points corresponding to the lower values of packing fractions ($\varphi$) were collected at the initial stage of compression. Since the granular material was placed in the cell without any prior preparation (such as tapping or mechanical vibration), the system initially existed in a low-density or disordered state. The decrease in volume induced a rearrangement of the grains, facilitating a transition toward a more densely packed phase. In the low-packing-fraction regime, the upper electrode encountered negligible mechanical impedance during the initial compression phase. Beyond a critical packing fraction (indicated by arrows in the corresponding plots of $\varphi$), the system undergoes frictional interlocking between constituent granules. This transition marks the onset of significant mechanical resistance to further volumetric strain. Upon completion of the compression cycle, the powder reaches a higher-density phase.

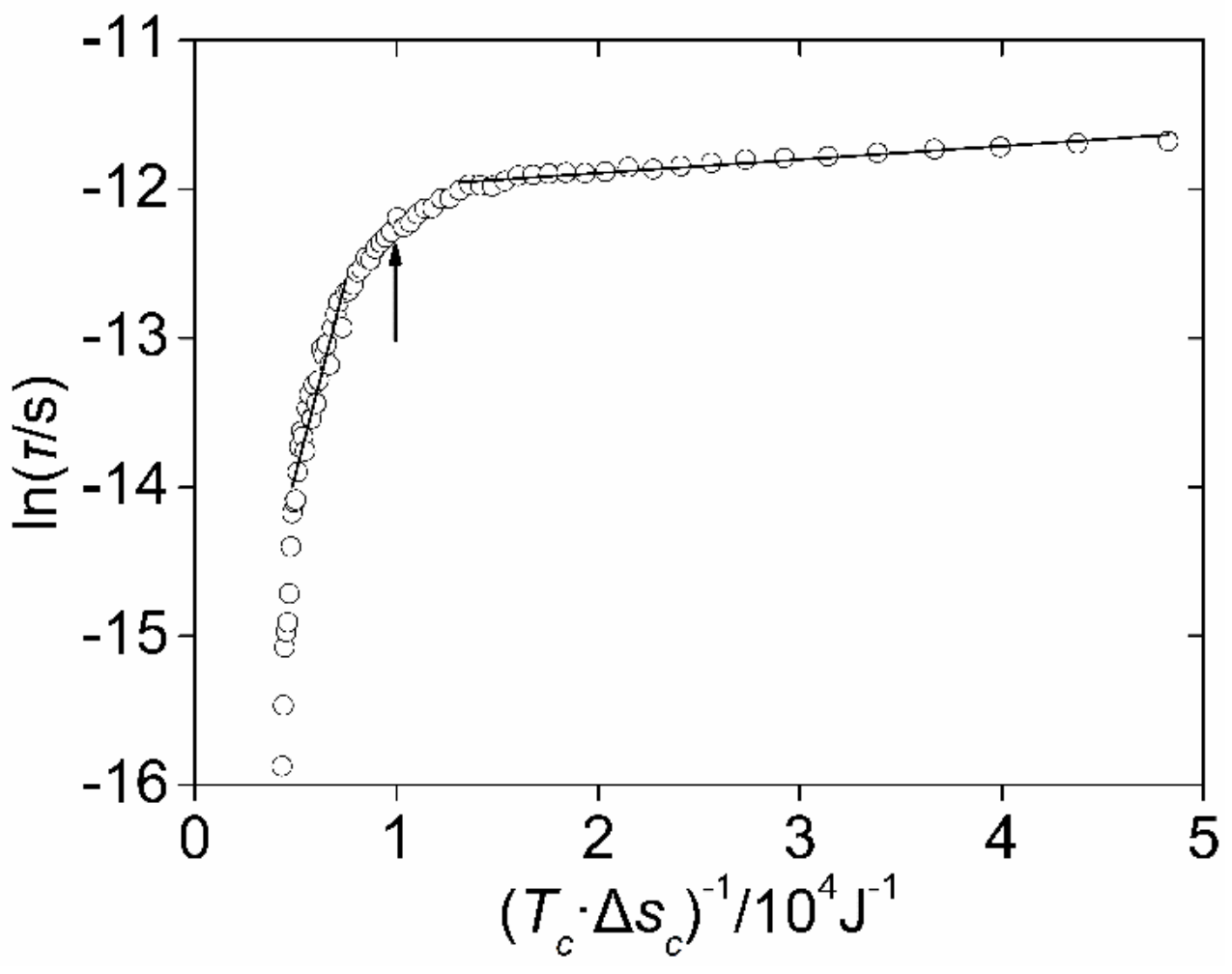

**Figure 6:** $ln\,\tau$ as a function of $(T_c \Delta s_c)^{-1}$ and the linear fit (solid line). The packing fraction increases from left to right.

The $ln\tau$ versus $[(T_c \Delta s_c)^{-1}]$ plot exhibits two linear regions, each spanning on side of the critical point, indicated by the arrow that signifies a transition from a loose to a dense state of granular matter. The scaling of $\ln \tau$ with $\left(\frac{1}{T_c \Delta s_c}\right)$ observed in both the low-$\varphi$ loose phase and the high-$\varphi$ dense phase is consistent with a modified AG relationship, $\tau \propto \exp\left(\frac{1}{T_c \Delta s_c}\right)$, which describes changes in the dielectric relaxation time $\tau$ as a function of $T_c \Delta s_c$ within each granular state. The slope of an individual linear trend (Figure 6) reflects the sensitivity of the overall electrical response of the granular system to mechanical stimulation, such as compression, vibration, or tapping. The change in slope near the loose-to-dense transition is likely associated with changes in frictional interlocking, contact topology, and the electrical percolation network, or with a combination of these effects. Moreover, structural changes may be associated with change in the slope of $T_c \Delta s_c$: distinct consecutive linear regions in a $ln\tau$ versus $(T_c \Delta s_c)^{-1}$ plot may indicate a transition near a critical packing fraction $\varphi_{c.}$, or equivalently near a critical value of $T_c(\varphi_{c.})\Delta s_c(\varphi_{c.})$. The existence of two linear regimes separated near the loose-to-dense transition is likely associated with the frictional interlocking, changes in coordination topology, changes in the electrical percolation network, or some combination of these effects. The observed configurational changes are accompanied by corresponding changes in the dielectric relaxation dynamics. Consequently, the observed scaling suggests a close coupling between the dielectric response and structural relaxation dynamics, although the latter are not independently measured in the present study.

## 5 Conclusion

In this study, we investigated the compaction dynamics of granular graphite by systematically varying its packing fraction within a cylindrical piston apparatus. Through concurrent volumetric and complex impedance measurements, we evaluated the system within the framework of Edwards' statistical thermodynamics for athermal systems, extracting both granular temperature and configurational entropy. Simultaneously, Broadband Dielectric Spectroscopy (BDS) was employed to measure dc conductivity and capacitance, allowing for the derivation of a characteristic dielectric relaxation time.

Drawing upon the Adam-Gibbs (AG) model, which traditionally relates structural relaxation time with temperature and configurational entropy in glass-forming liquids, we introduced a modified AG framework. By substituting classical thermal variables with their granular equivalents, we applied the modified AG framework to athermal granular systems investigated here. Applied to the gradually compressed graphite powder, this modified framework reveals a scaling relationship between the dielectric response and the structural relaxation dynamics. . This relationship indicates that variations in relaxation dynamics within the graphite are fundamentally governed by changes in configurational entropy.

Our findings support the applicability of a modified AG framework to the athermal systems investigated here. Furthermore, this study demonstrates the potential of dielectric measurements as a non-destructive approach for probing configurational changes and possible phase transitions during the compaction of granular matter.

**Research ethics:** Not applicable.

**Informed consent:** Not applicable.

**Author contributions:** **All** authors have accepted responsibility for the entire content of this manuscript and approved its submission. S.K. contributed data curation, formal analysis, investigation, methodology, validation, resources, visualization, writing – original draft, writing – review & editing. A.P. contributed conceptualization, methodology, resources, writing – review & editing, investigation, supervision.

**Use of Large Language Models, AI, and Machine Learning Tools:** None declared.

**Conflict of interest:** All other authors state no conflict of interest.

**Research funding:** None declared.

**Data availability:** Data is available on request from the authors.

---